\documentclass[aps,prl,twocolumn,10pt]{revtex4-2}

\usepackage[utf8]{inputenc}  
\usepackage[T1]{fontenc}
\usepackage[english]{babel}        
\usepackage{amsmath,amssymb,amsfonts,amsthm}
\usepackage{graphicx}
\usepackage{booktabs,array,multirow}

\usepackage[breaklinks=true,colorlinks=true,citecolor=blue,
filecolor=black,linkcolor=red,urlcolor=blue]{hyperref}

\begin{document}
	
\title{Generalized one-dimensional nonpolynomial Schr\"odinger equation for Bose-Einstein condensates with generic transverse confinement}

\author{Andr\'eia M. Basso}
\affiliation{Instituto de F\'isica,Universidade Federal de Goi\'as, 74.690-000 ,Goi\^ania, Goi\'as, Brazil}

\author{Wesley B. Cardoso}
\affiliation{Instituto de F\'isica,Universidade Federal de Goi\'as, 74.690-000 ,Goi\^ania, Goi\'as, Brazil}

\begin{abstract}
This work presents a dimensional reduction of Bose-Einstein condensates confined by generalized transverse potentials, parametrized by an exponent $n$. Starting from the three-dimensional Gross-Pitaevskii equation, we employ a variational \textit{ansatz} to derive an effective one-dimensional nonpolynomial Schr\"odinger equation (1D-NPSE), which self-consistently determines the transverse width dynamics. The model generalizes existing formalisms for cigar- and funnel-shaped geometries. We validate the approach through comprehensive numerical tests, demonstrating excellent agreement with full 3D simulations for ground-state properties across various interaction regimes. Finally, real-time simulations of matter-wave scattering at potential barriers verify the model's dynamical robustness, successfully replicating the spatiotemporal evolution and energy-dependent transmission characteristics observed in full 3D calculations.
\end{abstract}	
	
\maketitle

\section{Introduction}

In 1924, extending the quantum statistics formalism introduced by Bose \cite{Bose_JAA94}, Einstein \cite{Einstein_1924} predicted that a gas of non-interacting bosons of mass (m) exhibits a critical temperature below which a macroscopic population collapses into the ground state. This collective quantum phase transition, known as Bose–Einstein condensation (BEC), arises from the saturation of excited states at sufficiently low temperatures \cite{pitaevskiibose,Pethick}.

The first experimental realization of BEC occurred in 1995, when E. Cornell and C. Wieman condensed dilute vapors of ($^{87}\mathrm{Rb}$) \cite{Anderson_S95}, followed shortly thereafter by W. Ketterle with ($^{23}\mathrm{Na}$) \cite{Davis_PRL95} and R. G. Hulet in 1997 with ($^{7}\mathrm{Li}$) \cite{Bradley_PRL95}. These breakthroughs were made possible through advances in laser cooling techniques \cite{Hansch_OC75,Chu_PRL85} and the development of the magneto-optical trap \cite{Raab_PRL87}. Since then, BECs have become a powerful platform for probing macroscopic quantum phenomena, including bright \cite{Zakharov_JETP70, Strecker_N02} and dark \cite{Tsuzuki_JLTP71,Burger_PRL99,Denschlag_S00} solitons, molecular condensation \cite{Jochim_S03}, quantized vortices \cite{Matthews_PRL99,Madison_PRL00}, Anderson localization \cite{Billy_N08, Roati_N08}, interaction control via Feshbach resonances \cite{Cardoso_PRL10,Cardoso_PRL10_1,Cardoso_NA10,Serkin_PR10,Avelar_PR10,Cardoso_PRL10_2,Cardoso_PR12,Calaca_CNSNS14,Cardoso_CNSNS17,Calaca_OQE17,Calaca_EPJST18,Saravanan_CNSNS19,Cardoso_ND21,Cardoso_BJP21}, soliton train dynamics \cite{Strecker_N02}, and systems with spin–orbit coupling \cite{Lin_N11}, among others.

On the theoretical side, Bogoliubov’s formulation established a microscopic description for weakly interacting Bose gases \cite{Bogoliubov_JP47}, providing the foundation for mean-field theory. Within this framework, condensate dynamics are governed by the three-dimensional Gross–Pitaevskii equation (3D-GPE) \cite{Gross_NC61,Pitaevskii_SP61}, a nonlinear Schr\"odinger equation that incorporates both the external confinement and effective short-range interactions:
\begin{equation}
i \hbar \frac{\partial \psi}{\partial t} = \left( - \frac{\hbar^2}{2 m} \nabla^2 + V(\mathbf{r})+ \mathcal{N} g |\psi |^2\right) \psi , \label{gpe}
\end{equation}
where $\psi(\mathbf{r},t)$ is the macroscopic wave function normalized to unity, $\mathcal{N}$ the total number of particles, and $V(\mathbf{r})$ the external potential. The interaction strength is given by $
g = {2 \pi \hbar^2 a_s}/{m}$, with $a_s$ being the $s$-wave scattering length, positive (negative) for repulsive (attractive) two-body interactions.

For condensates confined in strongly anisotropic traps, where one or two spatial directions are tightly restricted, the dynamics effectively reduce to quasi-one- or quasi-two-dimensional regimes. Over the past decades, numerous theoretical approaches have been proposed to perform dimensional reduction of the GPE \cite{Jackson_PR98,Salasnich_PR02,Salasnich_SCM03,Zhang_PR05,Salasnich_PR06,Adhikari_PR06,Salasnich_PRA07,Salasnich_PRA07-ol,Salasnich_PR08,Maluckov_PR08,Adhikari_PR08,Mateo_PR08,Mateo_AP09,Salasnich_JP09,Salasnich_PR09,Adhikari_LPL09,Adhikari_IOP09,
Buitrago_JP09,Young_PR10,Cardoso_PR11,Salasnich_PR13,Young_PR13,Salasnich_PRA14,
Salasnich_OQE17,Couto_AP18,dos_Santos_PLPL19,dos_Santos_JP19,dos_Santos_PR21,Couto_CNSNS25}, aiming to accurately capture the effective low-dimensional dynamics of condensates in anisotropic geometries while retaining essential physical effects. In this sense, by using a variational approach and assuming cylindrical symmetry, in Ref. \cite{Salasnich_PR02} was derived an effective one-dimensional nonpolynomial Schr\"odinger equation for harmonically confined condensates. 

It is worth emphasizing that other studies have employed a formalism based on the adiabatic approximation, in which the transverse and longitudinal modes decouple \cite{Jackson_PR98,Mateo_PR08,Mateo_AP09,Couto_AP18,Couto_CNSNS25}. In this framework, the transverse width adapts instantaneously to its ground-state configuration, while the axial mode evolves independently. This approach leads to an effective one-dimensional equation that accurately describes the dynamics of repulsively interacting condensates.

Beyond these formulations, dimensional reduction methods have been generalized to diverse confinement geometries and interaction regimes. Examples include \textit{tube}- \cite{dos_Santos_PLPL19}, \textit{funnel}- \cite{dos_Santos_JP19}, pancake- and disk-shaped condensates \cite{Salasnich_PR09,Mateo_PR08}, periodic transverse potentials \cite{Salasnich_PRA07-ol}, and systems with spatially modulated nonlinearities \cite{Young_PR13}. Further extensions address anisotropic harmonic traps \cite{Salasnich_JP09}, elongated traps with vortex dynamics \cite{Mateo_AP09,Salasnich_SCM03}, binary mixtures \cite{Young_PR10}, spinor condensates \cite{Zhang_PR05}, and multicomponent systems with spin–orbit and Rabi couplings \cite{Salasnich_PR13,Salasnich_PRA14}. Applications also include vector solitons \cite{Salasnich_PR06}, discrete lattices \cite{Maluckov_PR08}, Anderson localization in binary gases \cite{dos_Santos_PR21}, as well as fermionic mixtures and superfluid systems \cite{Buitrago_JP09,Adhikari_PR06,Adhikari_LPL09,Adhikari_IOP09,Adhikari_PR08}.

In this work, we employ a variational dimensional reduction method to derive effective equations for generalized transverse confinements, extending the standard \textit{cigar}- \cite{Salasnich_PR02} and \textit{funnel}-shaped \cite{dos_Santos_JP19} models. By introducing a transverse potential parameterized by a positive real exponent $n$, we analyze both stationary and dynamical properties of the resulting effective one-dimensional equations for different integer values of $n$, providing a unified framework for condensates in generalized confinement geometries.

\section{Generalized effective nonpolynomial Schr\"odinger equation}

We consider a Bose-Einstein condensate subject to a separable trapping potential of the form  
\begin{equation}
V(\mathbf{r}) = V_\perp(r) + V_\parallel(z),
\end{equation}
where $V_\parallel(z)$ denotes an arbitrary axial potential and $V_\perp(r)$ describes the transverse confinement with $r=\sqrt{x^2+y^2}$. The transverse potential is introduced in the generalized nonharmonic form  
\begin{equation}
V_\perp(r) = \frac{m \omega_\perp^2 n^2}{8} 
\left( \frac{\alpha^{4-2n} r^{2n} - 2 \beta^{4-n} r^n}{r^2} \right),
\label{potgeralcharuto}
\end{equation}
where $\omega_\perp$ is the transverse frequency, $\alpha$ and $\beta$ are characteristic length scales, and $n \in \mathbb{R}^{*}_{+}$ is a  parameter.

\begin{figure}[tb]
\centering \includegraphics[width=\columnwidth]{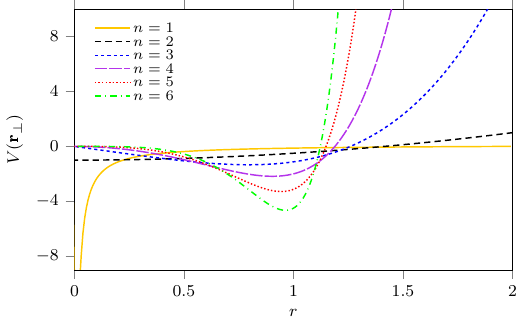} 

\caption{Transverse potential defined by Eq.~(\ref{potgeralcharuto}), expressed in units of $m\omega_{\perp}^2$, as a function of the radial coordinate $r$ for $n=1$, $2$, $3$, $4$, $5$, and $6$. The parameters are fixed at $\alpha=1$ and $\beta=1$.}

\label{fig:potcharutogeralnovo}
\end{figure}

For $n<0$, the potential vanishes asymptotically as $r \to \infty$ while diverging positively as $r \to 0$, which implies the presence of a hard barrier at the origin and a decaying tail at large distances, i.e., the confinement is not strict at infinity. When $n>1$, the potential may exhibit a local minimum located at  
\begin{equation}
r_c = \left( \frac{n-2}{n-1} \right)^{1/n}.
\end{equation}
It follows that for $n=1$ the potential has no minimum and diverges monotonically as $r \to 0$, whereas for $n >1$ the minimum shifts with increasing $n$ (with $\lim _{n\rightarrow\infty} \{r_c\}=1$), resulting in a narrower confining region around $r_c$ and a sharper growth for $r>r_c$ (see Fig.~\ref{fig:potcharutogeralnovo}).  

The purpose of this construction is to derive a family of generalized effective one-dimensional nonpolynomial Schr\"odinger equations governing cigar-shaped condensates confined by potentials of the type given by Eq.~(\ref{potgeralcharuto}). Note that the cases $n=1$ and $n=2$ correspond to those previously investigated in Refs. \cite{dos_Santos_JP19} and \cite{Salasnich_PR02}, respectively.

Substituting Eq.~(\ref{potgeralcharuto}) into the 3D-GPE (\ref{gpe}) and performing the rescaling 
$(x,y,z) \rightarrow (x,y,z)/a_\perp$, $t \rightarrow \omega_\perp t$, $\psi \rightarrow a_\perp^{3/2}\psi$, 
$g \rightarrow \frac{2\mathcal{N}a_s}{a_\perp}$, $(\alpha,\beta)\rightarrow (\alpha,\beta)/a_\perp$,
with $a_\perp=\sqrt{\hbar/(m\omega_\perp)}$ being the transverse length scale, and setting $(\alpha,\beta)=1$, for simplicity, we obtain the dimensionless 3D-GPE:
\begin{equation}
i \frac{\partial \psi}{\partial t} = -\tfrac{1}{2}\nabla^2\psi 
+ \tfrac{n^2}{8}\left(\frac{r^{2n}-2r^n}{r^2}\right)\psi 
+ V(z)\psi + 2\pi g|\psi|^2\psi. 
\label{charuto3dn}
\end{equation}
The corresponding Lagrangian density reads
\begin{eqnarray}
\mathcal{L}_{3D} &=& \tfrac{i}{2}\left(\psi\frac{\partial \psi^*}{\partial t}-\psi^*\frac{\partial \psi}{\partial t}\right)
+ \tfrac{1}{2}|\nabla \psi|^2 \nonumber \\
&+& \left[\tfrac{n^2}{8}\left(\frac{r^{2n}-2r^n}{r^2}\right)+V(z)\right]|\psi|^2 
+ \pi g|\psi|^4.
\label{lagcharutogeral}
\end{eqnarray}

To derive an effective one-dimensional model, we apply a variational approach using a generalized \textit{ansatz}, given by 
\begin{equation}
\psi(\mathbf{r},t) = 
\sqrt{\tfrac{n}{2\pi \Gamma(2/n)}} 
\exp\!\left(-\tfrac{r^n}{2\sigma^2}\right) 
\frac{f(z,t)}{\sigma^{2/n}},
\label{ansatzcahrutogeral}
\end{equation}
with $\Gamma$ being the Euler Gamma function, $\sigma(z,t)$ is the transverse width and $f(z,t)$ the normalized axial wavefunction.  
This construction guarantees the reduction of the 3D density to its effective 1D counterpart:
\begin{equation}
2\pi \int_0^\infty |\psi(r,\theta,z)|^2 r\,dr = |f(z,t)|^2.
\end{equation}

Substituting the \textit{ansatz} from Eq.~(\ref{ansatzcahrutogeral}) into the Lagrangian density (\ref{lagcharutogeral}) and integrating over the transverse coordinates yields an effective one-dimensional Lagrangian. This derivation employs an adiabatic approximation \cite{Mateo_AP09, Jackson_PR98, Salasnich_PR02, Salasnich_PR02_1}, which posits a separation of time scales between the fast radial and slow axial dynamics. Consequently, the transverse width is treated as constant, with the radial density profile adapting instantaneously to the axial dynamics.
Consequently, we derive the effective one-dimensional Lagrangian, given by
\begin{eqnarray}
	\mathcal{L}_{eff} &=& \frac{i}{2}\left(f \frac{\partial f^*}{\partial t} -f^* \frac{\partial f}{\partial t}   \right) + \frac{1}{2} \left | \frac{\partial f }{\partial z}\right|^2 + [V(z)+1] |f|^2 \nonumber \\ &+&
 \frac{n^2\left( \sigma^2 -1\right)^2}{8\; \sigma^{4/n} \;\Gamma\left(\frac{2}{n}\right)} |f|^2 + \frac{ g \;n\;}{\sigma^{4/n}\; 2^{\frac{n+2}{n}}\;\Gamma\left(\frac{2}{n}\right)} |f|^4. \label{efetivacharutogeral}
\end{eqnarray}

The variation of this Lagrangian density with respect to the fields $f^*$ and $\sigma$ yields the corresponding Euler-Lagrange equations:
\begin{eqnarray}
	i \frac{ \partial f}{\partial t} &=& -\frac{1}{2} \frac{\partial^2 f}{\partial z^2}+  V(z) f \nonumber \\ &+&\frac{n}{\sigma^{4/n}\; \Gamma\left( \frac{2}{n}\right)}\left[ \frac{n}{8} \left( \sigma^2 -1\right)^2+\frac{g}{2^{2/n}} |f|^2\right]f \label{efetivancharuto},\\
	\sigma^2_{\pm} &=& \frac{ n - 2 \pm \sqrt{4^{\frac{2n-1}{n}}g |f|^2\frac{(n-1)}{n}+n^2}}{2  (n-1)} \label{sigmancharutogeral},
\end{eqnarray}
where the signs in $\sigma^2_{\pm}$ denote the two possible solutions for $\sigma^2$, which will be examined in detail subsequently.

\begin{figure}[tb]
\centering
\includegraphics[width=\columnwidth]{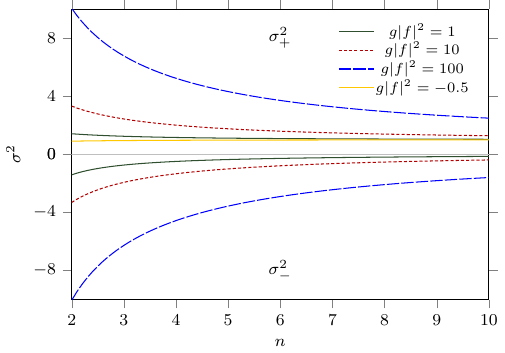}
\caption{ Solutions for the variational field \( \sigma^2_+ \) and \( \sigma^2_- \) in the upper and lower quadrants, respectively, as a function of \( n \) for different values of \( g|f|^2 \).}
\label{fig:sigmac}
\end{figure}

Next, the effective chemical potential is derived by considering stationary solutions ($f(z,t) = f(z)e^{-i\mu t}$). Inserting this \textit{ansatz} into Eq.~(\ref{efetivancharuto}) and enforcing the normalization condition provides:
\begin{eqnarray}
	\mu &=& \int dz\left\{ \frac{1}{2}\left|\frac{\partial f}{\partial z}\right|^2+ V(z)|f|^2 \right.  \nonumber \\  &+& \left. \frac{n}{\sigma^{4/n}\; \Gamma\left( \frac{2}{n}\right)}\left[ \frac{n}{8}\left( \sigma^2 -1\right)^2 + \frac{g}{2^{2/n}} |f|^2\right]|f|^2 \right\}. 
\label{potefetivo}
\end{eqnarray}
Figure~\ref{fig:sigmac} shows that the $\sigma_{-}^2$ branch yields negative field values, resulting in a complex-valued chemical potential. This behavior arises from the last term in Eq.~(\ref{potefetivo}), which depends on $\sigma^{4/n}$ and leads to unphysical solutions. Consequently, the physical solution is given by the $\sigma_{+}^2$ branch, which will be adopted throughout this work.

Substituting $n=1$ into the potential defined by Eqs.~(\ref{potgeralcharuto}) and ~(\ref{efetivancharuto}) yields:
 \begin{equation}
 	V_\perp(r) = \frac{1}{4}\left(\frac{1}{2}-\frac{1}{r}\right),
 \end{equation}
 and the effective equations take the form:
 \begin{eqnarray}
 	i \frac{ \partial f}{\partial t} &=& -\frac{1}{2} \frac{\partial^2 f}{\partial z^2}+ V(z)f \nonumber \\ &+&\frac{1}{8}\left[1+\frac{1}{\sigma^2}\left(\frac{1}{\sigma^2} - 2\right) \right]f +\frac{g}{4} \frac{|f|^2}{\sigma^4}f,
 \end{eqnarray}
with $\sigma^2 = 1+g|f|^2$. These equations are consistent with those derived in \cite{dos_Santos_JP19} for a funnel-shaped condensate, which is characterized by a potential singularity at $r=0$. The results differ only by rescaling factors and additive constants, which are dynamically irrelevant as they do not affect the equations of motion.
 
The application of this formalism to the case $n = 2$ results in the following equations: 
\begin{eqnarray}
V_\perp(r) &=& \frac{1}{2}\left(r^2 - 1 \right),\nonumber \\
i \frac{ \partial f}{\partial t} &=& -\frac{1}{2} \frac{\partial^2 f}{\partial z^2}+ \gamma_C^2 \frac{z^2}{2} f \nonumber \\ &+& \left[\frac{1}{2} \left(\frac{1}{\sigma^2}+\sigma^2\right)-1 \right]f +\frac{g}{\sigma^2}|f|^2f,
\end{eqnarray}
with $\sigma^4 = 1+g|f|^2$, which reproduces the results of \cite{Salasnich_PR02} for a cigar-shaped condensate confined in a harmonic trap. In the above expression, $\gamma_c = \omega_z/\omega_\perp \ll 1$ defines the anisotropy parameter of the trap.
 
Consequently, the selection of a suitable \textit{ansatz} yields the one-dimensional equations (\ref{efetivancharuto}), hereafter designated as the 1D-EFG model, alongside the analytical expressions for the variational fields (\ref{sigmancharutogeral}) that govern the transverse width. This formulation generalizes previously reported results for both cigar- and funnel-shaped condensates \cite{Salasnich_PR02,dos_Santos_JP19}. The resulting generalized effective equations are valid for a continuous range of the transverse confinement parameter, $n \in \mathbb{R}_{+}^{*}$, establishing a unified framework for comparing diverse systems. For a comparative analysis, the following section examines different integer values of $n$ within this generalized model.

\subsection*{Cubic Equation}

We also consider a comparative effective 1D cubic model, derived from the 3D-GPE (Eq.~\ref{charuto3dn}) via a variational \textit{ansatz} that neglects transverse compressibility, given by
\begin{equation}
	\psi(\textbf{r}, t)= \sqrt{\frac{n}{2\pi \Gamma\left( \frac{2}{n}\right)}} \cdot e ^{- r^n} f ,
\end{equation}
which applying the same variational procedure outlined above yields the corresponding effective one-dimensional equation with cubic nonlinearity (1D-CN):
\begin{equation}
	i \frac{\partial f }{\partial t} = - \frac{1}{2} \frac{\partial ^2 f}{\partial z^2} + V(z) f + \frac{g n}{\Gamma \left( \frac{2}{n}\right)} |f|^2 f. \label{reduzida}
\end{equation}
Figure~\ref{fig:sigmac} demonstrates that $\sigma^2$ asymptotically approaches unity as $n \rightarrow \infty$, with the convergence rate inversely proportional to the magnitude of $g|f|^2$. This behavior is corroborated by Fig.~\ref{fig:potcharutogeralnovo}, where an increasingly narrow transverse confinement region is observed for larger $n$, indicating a diminished role of the variational field $\sigma$. 

\section{Numerical Results}

The effective reduced-dimensional formalism described by Eqs. (\ref{efetivancharuto}) and (\ref{sigmancharutogeral}) lacks closed-form analytical solutions. To conduct a comparative analysis of the solutions across the parameter $n$, we perform a series of imaginary and real time numerical simulations based on the second-order \textit{Split-Step} method \cite{yang}.

The formalism is exemplified using the following axial potential
\begin{equation}
V(z) = \gamma^2 \frac{z^2}{2},
\end{equation}
where $\gamma = \omega_z/\omega_\perp \ll 1$ denotes the anisotropy parameter of the system, with $\omega_z$ accounting for the longitudinal frequency, which controls the strength of the transverse confinement. In this context, we set $\gamma^2 = 0.1$.

\begin{figure}[tb]
\centering
\includegraphics[width=\columnwidth]{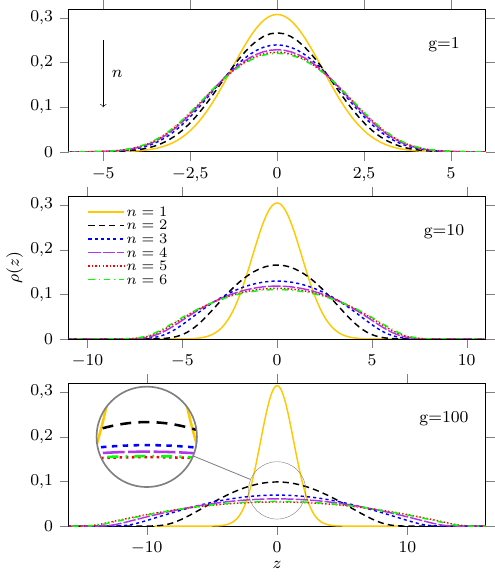}
\caption{ Axial density for the effective 1D-EFG  (\ref{efetivancharuto}) for different values of the interaction strength, i.e., $g = 1$, $10$, and $100$.}
\label{fig:densidadecharutogeral}
\end{figure}

We first compare the  axial density profile obtained from the effective equation (\ref{efetivancharuto}), given by $\rho(z)_{1D} = |f(z,t)|^2$, for the first six integer values of $n$ and different values of $g$. As depicted in Fig.~\ref{fig:densidadecharutogeral}, increasing the transverse confinement parameter and the interaction strength induces a significant axial elongation of the density profile.

To assess the accuracy of the effective description, we compare the maximum value of the axial density predicted by the one-dimensional equations (\ref{efetivancharuto}) and (\ref{reduzida}) with the corresponding density profile obtained from the full three-dimensional GPE (\ref{charuto3dn}):
\begin{equation}
	\rho(z)_{3D} = 2 \pi \int |\psi(\textbf{r},t)|^2 r dr.
\end{equation}

\begin{figure}[tb]
\centering
\includegraphics[width=\columnwidth]{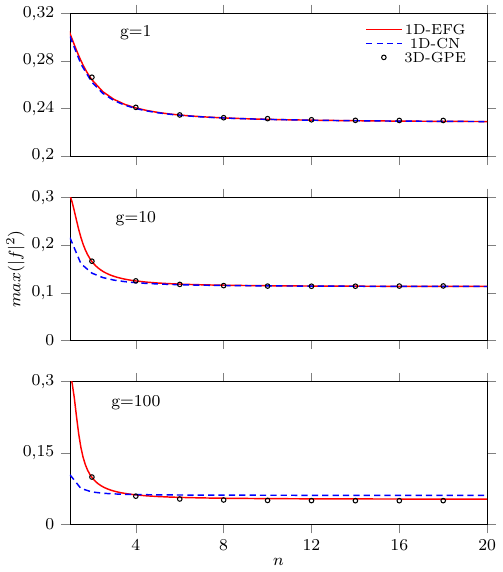}
\caption{Maximum value of the axial density as a function of $n$ for the 1D-EFG (\ref{efetivacharutogeral}),  the 1D-CN (\ref{reduzida}) and the 3D-GPE~(\ref{charuto3dn}), for different values of the interaction strength, i.e., $g = 1$, $10$, and $100$.}
\label{fig:maxu}
\end{figure}

Figure~\ref{fig:maxu} demonstrates that the generalized effective model exhibits excellent agreement with the full three-dimensional solution. In contrast, the cubic model displays increasingly significant deviations for stronger interatomic interactions.

We further characterize the system by examining the effective chemical potentials and the expectation values of the squared longitudinal length, 
the latter being defined respectively for the reduced and full models as
\begin{eqnarray}
\langle z^2\rangle_{1D} = \int z^2 |f|^2 dz,\\
\langle z^2\rangle_{3D} = \int z^2 |\psi|^2 d\textbf{r}.
\end{eqnarray} 

\begin{figure}[tb]
\centering
\includegraphics[width=\columnwidth]{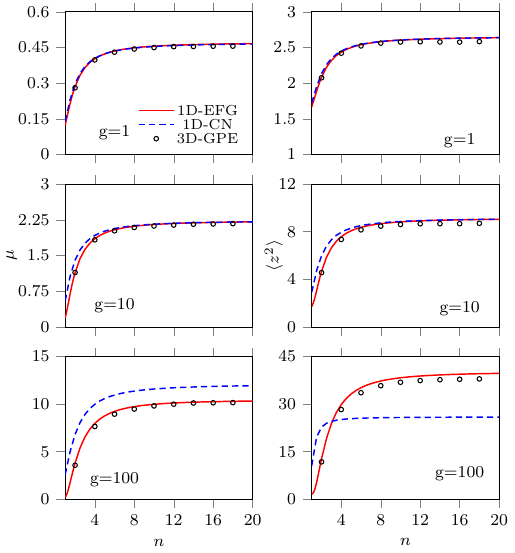}
\caption{Chemical potential (left) and expectation value of the squared longitudinal length (right) for the 1D-EFG (\ref{efetivacharutogeral}),  the 1D-CN (\ref{reduzida}) and the 3D-GPE~(\ref{charuto3dn}), as a function of $n$ for different values of the interaction strength, \textit{viz}., $g = 1$, $10$, and $100$.}
\label{fig:potquimico}
\end{figure}

Figure~\ref{fig:potquimico} reveals that as the transverse confinement is strengthened (increasing $n$), both quantities converge toward a linear dependence, with the most pronounced nonlinear variations occurring at lower $n$. Furthermore, the accuracy of the generalized effective model (\ref{efetivancharuto}) deteriorates with increasing scattering amplitude $g$. This discrepancy arises because stronger repulsive interactions enhance the coupling between transverse and axial degrees of freedom, thereby violating the weak-coupling assumption inherent to the factorized \textit{ansatz} upon which the dimensional reduction is based.

The split-step Fourier method, when applied in imaginary time, serves as a powerful computational tool to probe the linear stability of localized solutions in the nonlinear Schr\"odinger equation. This formulation effectively transforms the equation into a dissipative gradient flow, guiding the wave function toward the system's ground state. The algorithm's iterative application of the nonlinear and kinetic energy operators provides a discrete approximation to this continuous minimization process. Crucially, the numerical stability of this propagation is dictated by the Bogoliubov-de Gennes spectrum of the stationary solution. The presence of an eigenmode with a positive eigenvalue, indicating a dynamical instability in real time, manifests as an exponential numerical divergence during imaginary-time evolution. This occurs because the imaginary-time propagator amplifies such unstable modes. Therefore, the robust and monotonic convergence of the split-step algorithm in imaginary time not only yields the stationary profile but also provides strong empirical evidence for its linear stability, confirming the absence of negative curvature in the energy landscape \cite{yang}.

Consequently, the imaginary-time convergence criterion provides a robust numerical basis for identifying the critical interaction parameter $g_c$. Systematic simulations of the axial density profile, chemical potential, and squared longitudinal length reveal characteristic discontinuities at the stability threshold, precisely delineating the parameter region $g<g_c$ where no stable stationary solutions exist.

\begin{table}[tb]

		\centering
		\begin{tabular}{|l|c|c|}
			\toprule[2pt]
			$\textbf{n}$  & \textbf{3D-GPE} & \textbf{1D-EFG}  \\
			\cmidrule(r){1-1} \cmidrule(lr){2-3} 
			$\textbf{1}$  & $-0.85$ & $-0.95$ \\ 
			$\textbf{2}$  & $-1.28$ & $-1.25$\\ 
			$\textbf{3}$  & $-1.39$ & $-1.50$\\ 
			$\textbf{4}$ & $-1.47$  & $-1.85$ \\ 
			$\textbf{5}$  & $-1.53$ & $-2.18$ \\ 
			$\textbf{6}$  & $-1.58$ & $-2.51$ \\ 
			\bottomrule[1.5pt]
		\end{tabular}
	\caption{Values of the critical interaction strength $g_c$ computed from the full 3D Gross-Pitaevskii equation (\ref{charuto3dn}) and the effective 1D-EFG model (\ref{efetivacharutogeral}) for a range of transverse confinement exponents $n = 1$, $2$, ..., $6$.}

	\label{tabela:gc}
\end{table}

The critical interaction parameters $g_c$ for each model are summarized in Table~\ref{tabela:gc}. The quantitative differences between the values predicted by the effective one-dimensional equations and the full three-dimensional calculation are inherent to the approximations involved in the dimensional reduction. Despite this, the 1D-EFG model (\ref{efetivancharuto}) correctly captures the essential physics, as evidenced by their consistency with the full 3D results for the axial density, chemical potential, and the expectation value of the squared longitudinal length. In contrast, the standard one-dimensional cubic nonlinear equation (\ref{reduzida}) fails to predict the collapse phenomenon altogether.

\subsection{Dynamics of wave packet-barrier scattering}

The dynamical behavior of the effective model was further investigated by simulating the scattering of a wave packet from a Gaussian potential barrier centered at $z=0$ \cite{Nguyen_N14, Manju_PR18}. The initial state $f(z)$ was prepared via imaginary-time propagation, using a displaced harmonic axial potential, given by
\begin{equation}
	V(z) = \frac{(z - z_0)^2}{2}.
\end{equation}
The dynamics are initiated at $t=0$ by the sudden removal of the external potential, concurrent with the application of a momentum impulse $p$ that accelerates the wave packet towards the barrier, i.e., 
\begin{equation}
	f(z) \rightarrow f(z)\exp(ipz).
\end{equation}
The scattering potential is defined by a Gaussian barrier centered at $z=0$, given by
\begin{equation}
	V_b(z) = A\exp\left(-\frac{z^2}{B^2}\right), 
	\label{potbarreira}
\end{equation}
where $A$ is the peak amplitude and $B$ characterizes the width of the barrier.

\begin{figure}[tb]
	\centering
	\includegraphics[width=\columnwidth]{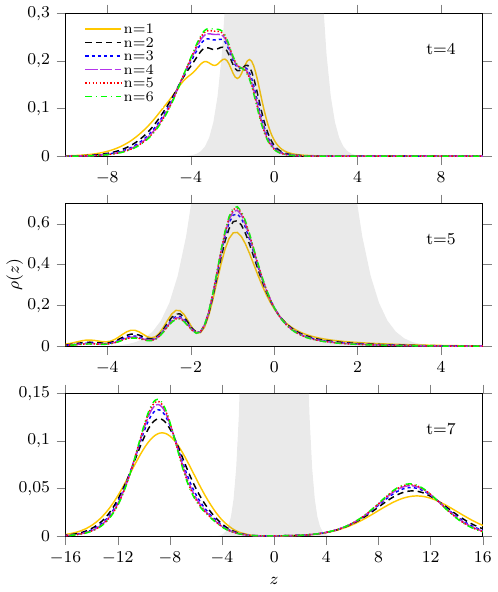}
	\caption{Axial density for the 1D-EFG (\ref{efetivancharuto}) over time, with initial momentum $p = 3$ and $g= -0.5$, colliding with a Gaussian potential barrier (\ref{potbarreira}) with parameters $A = 5$ and $B = \sqrt{2}$. The barrier profile is depicted in light gray.}
\label{fig:barreiracharutogeral}
\end{figure}

Real-time simulations of the scattering process, presented in Fig.~\ref{fig:barreiracharutogeral}, reveal the evolution of the axial density profile. The results reveal that the system displays qualitatively similar dynamics for all six distinct values of $n$, with increasingly pronounced interference effects with the potential barrier as this parameter decreases.

\begin{figure}[tb]
	\centering
	\includegraphics[width=\columnwidth]{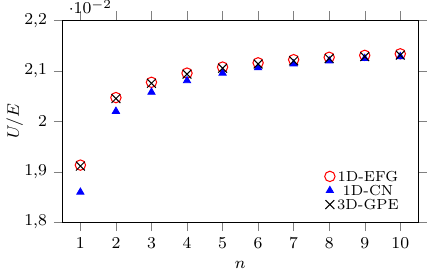}
	\caption{Ratio between kinetic and potential energies as a function of $n$ during the collision with a Gaussian potential barrier (\ref{potbarreira})  at $t = 10 s$, with parameters $A = 5$ and $B = \sqrt{2}$, interaction strength $g = -0.5$, and momentum $p = 3$. Results are shown for 1D-EFG (\ref{efetivacharutogeral}),  the 1D-CN (\ref{reduzida}) and the 3D-GPE~(\ref{charuto3dn}).}
	\label{fig:momento}
\end{figure}

To evaluate the performance of the generalized model in dynamical scenarios, we calculate the transmission coefficients for the full and effective models as
\begin{eqnarray}
	Tr_{3D} &=& 2 \pi \int _{0}^{\infty} dz \int_{0}^{\infty} r |\psi|^2 dr, \\
	Tr_{1D} &=& \int_{0}^{\infty}|f|^2 dz.
\end{eqnarray}
These quantities are computed at a time $t$ after the wave packet has completely passed through the center of scattering potential. Figure~\ref{fig:momento} demonstrates that the transmission probability increases monotonically with the confinement parameter $n$. Furthermore, the generalized model accurately captures the three-dimensional transmission characteristics, while the standard cubic model shows measurable deviations.

\begin{figure}[tb]
	\centering
	\includegraphics[width=\columnwidth]{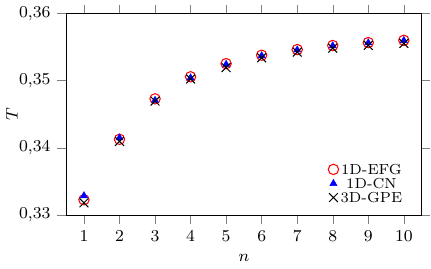}
\caption{Transmission coefficient as a function of $n$ during the collision with a Gaussian barrier (\ref{potbarreira})  at $t = 10 s$ with $A = 5$ and $B = \sqrt{2}$, adopting $g = -0.5$ and $p = 3$, for the 1D-EFG (\ref{efetivacharutogeral}),  the 1D-CN (\ref{reduzida}) and the 3D-GPE~(\ref{charuto3dn}).
}
\label{fig:transmissao}
\end{figure}

Finally, we analyze the ratio between the kinetic and interaction energies in the presence of the potential barrier. For the three-dimensional model, these energy contributions are defined as
\begin{eqnarray}
	U_{3D} &=& 2 \pi \int_{-\infty}^{+\infty}V_b \;dz \int ^{\infty}_{0}|\psi|^2 \;r \;dr, \\
	E_{3D} &=& \pi \int^{+ \infty}_{-\infty}\int_{0}^{\infty}\left(\left|\frac{\partial \psi}{\partial r}\right|^2+ \left|\frac{\partial \psi}{\partial z}\right|^2\right)r\;dr\;dz.
\end{eqnarray}
For the one-dimensional reductions, the corresponding expressions read
\begin{eqnarray}
	U_{1D} &=&\int_{-\infty}^{\infty}V_b |f|^2 dz \\
	E_{1D} &=& \int^{+ \infty}_{-\infty}\frac{1}{2}\left|\frac{\partial f}{\partial z}\right|^2
\end{eqnarray}

n Figs.~\ref{fig:transmissao} and \ref{fig:momento}, both the transmission coefficient and the kinetic-to-interaction energy ratio are found to increase with $n$, demonstrating the high accuracy of the 1D-EFG model (\ref{efetivancharuto}) relative to the full three-dimensional profile. In contrast, the 1D-CN model (\ref{reduzida}) exhibits noticeable discrepancies for small $n$, while its agreement improves as $n$ increases.

These observations are further quantified by the maximum relative deviations in the transmission coefficients. When compared with the full 3D-GPE (\ref{charuto3dn}), the 1D-EFG and 1D-CN formulations yield differences of $0.094\%$ and $0.153\%$ for $n=2$; $0.108\%$ and $0.168\%$ for $n=5$; and $0.119\%$ and $0.122\%$ for $n=10$, respectively.

\section{Conclusion}

This work introduces a generalized trapping potential (\ref{potgeralcharuto}) that unifies the cigar- \cite{Salasnich_PR02} and funnel-shaped \cite{dos_Santos_JP19} condensate geometries within a single theoretical framework. Through variational dimensional reduction, we have derived the corresponding effective one-dimensional equation (\ref{efetivancharuto}) and obtained analytical expressions for the variational fields governing the transverse width (\ref{sigmancharutogeral}). For comparison, we also present a simplified cubic model (\ref{reduzida}) derived by neglecting transverse compressibility.

The resulting formalism is valid for a continuous range of transverse confinement exponents, $n \in \mathbb{R}_{+}^{*}$, establishing a unified framework for analyzing confinement effects across different geometries. We have numerically validated this approach for selected integer values of $n$, demonstrating that the ground-state solutions obtained from our effective model show excellent agreement with full 3D-GPE simulations. This agreement extends to key stationary properties, including the maximum axial density, chemical potential, and expectation value of the squared axial length $\langle z^2 \rangle$, with our model consistently outperforming the standard cubic approximation. The critical scattering amplitudes for collapse were also accurately determined.

In dynamical scenarios involving scattering from a Gaussian potential barrier, the model correctly captures the transmission characteristics and energy redistribution during interaction. Quantitative analysis reveals a systematic increase in the reflected fraction with increasing $n$, consistent with the enhanced transverse confinement.

The principal advantage of this generalized approach lies in its ability to model realistic experimental conditions, where confinement potentials often deviate from ideal symmetric forms and may vary continuously. By encapsulating the transverse geometry through a single exponent $n$, our framework enables systematic exploration of confinement effects, regime transitions, and trap optimization within a unified mathematical structure.

In summary, this work provides a versatile and systematically controllable framework that significantly enhances the predictive capability and practical applicability of effective models for Bose-Einstein condensates under generalized confinement.

\subsection*{Acknowledgements}

This work was supported by the Brazilian funding agencies: Conselho Nacional de Desenvolvimento Cient\'ifico e Tecnol\'ogico (CNPq) [Grant No. 306105/2022-5], Coordena\c{c}\~ao de Aperfei\c{c}oamento de Pessoal de N\'ivel Superior (CAPES), and Funda\c{c}\~ao de Amparo à Pesquisa do Estado de Goi\'as (FAPEG). Additional support was provided by the Brazilian National Institute of Science and Technology for Quantum Information (INCT-IQ) [Grant No. 465469/2014-0].

	\bibliographystyle{apsrev4-2} 
	\bibliography{bib}        
	
\end{document}